%% file: NoCMC-incompleteness-Bartnik.tex
\documentclass[12pt]{amsart}
\usepackage{amsmath}
\usepackage{amssymb}
\usepackage{amsthm}
\usepackage{enumerate}
\usepackage{graphicx}
\usepackage{tikz-cd}
\usepackage{bbm}
\usepackage{setspace}
\usepackage{chngcntr}
\usepackage{hyperref}
\hypersetup{colorlinks=true, linkcolor=blue}

\newtheorem{theorem}{Theorem}
\newtheorem*{theorem*}{Theorem}
\newtheorem*{lemma*}{Lemma}
\newtheorem*{prop*}{Proposition}
\newtheorem*{cor*}{Corollary}
\newtheorem*{claim*}{Claim}

\newtheorem{lemma}[theorem]{Lemma}

\theoremstyle{definition}

\newtheorem{remark}{Remark}

\newtheorem*{definition*}{Definition}
\newtheorem*{remark*}{Remark}

\counterwithin{figure}{section}

\input{MyDefns}

\title[Null geodesic incompleteness of  spacetimes]
{Null geodesic incompleteness of spacetimes with no CMC Cauchy surfaces}\author{Madeleine Burkhart}
\address[Madeleine Burkhart]{University of Washington, Seattle, WA USA}
\email{burkhm2@uw.edu}
\author{Martin Lesourd}
\address[Martin Lesourd]{University of Oxford, Oxford, UK}
\email{martin.lesourd@linacre.ox.ac.uk}
\author{Daniel Pollack}
\address[Daniel Pollack]{University of Washington, Seattle, WA USA}
\email{pollack@uw.edu}

\begin{document}
\setcounter{section}{0}

\maketitle

\begin{abstract}
Chru{\'s}ciel, Isenberg, and Pollack constructed a class of vacuum cosmological spacetimes that do not admit Cauchy surfaces with constant mean curvature. We prove that, for sufficiently large values of the gluing parameter, these examples are  both future and past null geodesically incomplete.

The authors are honored to dedicate this paper to Robert Bartnik on the occasion of his 60th birthday.
\end{abstract}

\section{Introduction} \label{sec:intro}

It is well-known that constant mean curvature (CMC) Cauchy surfaces play an important role in the mathematical study of solutions to the Einstein field equations.  When solving the Einstein constraint equations via the conformal method, the CMC assumption ensures that  the resulting equations  semi-decouple, hence leading to a far more robust understanding of existence and uniqueness than in the general case. The CMC gauge is also quite useful for studying the Einstein evolution equations, both analytically and numerically. This vital role of CMC Cauchy surfaces raises an interesting open question: When do globally hyperbolic spacetimes admit CMC Cauchy surfaces? It is known that not all globally hyperbolic spacetimes have CMC Cauchy surfaces: In \cite{B88}, Bartnik found no-CMC spacetimes with dust, and in \cite{CIP05}, Chru{\'s}ciel, Isenberg, and Pollack (CIP) found a family of vacuum spacetimes with no CMC Cauchy surfaces using a modified form of \textit{IMP gluing}: a connected sum gluing procedure developed initially by Isenberg, Mazzeo and Pollack (see \cite{CD03}, \cite{C00}, \cite{IMaxP05}, \cite{IMP02}, \cite{IMP03}). We refer to the recent survey \cite{DH17} of Dilts and Holst for some results and conjectures related to this question.

In a recent paper \cite{GL18a}, Galloway and Ling proved a new existence result for CMC slices: every future timelike geodesically complete cosmological spacetime (recall that a \textit{cosmological spacetime} is a globally hyperbolic spacetime with compact Cauchy surfaces) with everywhere nonpositive timelike sectional curvatures must admit a CMC Cauchy surface. Motivated by the Bartnik splitting conjecture (see Conjecture 2 of \cite{B88}), the conjectures of \cite{DH17}, and the conditions of their own result, Galloway and Ling conjecture that a future timelike geodesically complete cosmological spacetime satisfying the strong energy condition must contain a CMC Cauchy surface. In light of this conjecture, geodesic completeness or incompleteness of no-CMC spacetimes becomes relevant. Noting that Bartnik's examples are by construction timelike geodesically incomplete to both the  future and the past, we turn our attention toward the CIP examples.

In this note, we prove the following:
\begin{theorem}
For sufficiently large gluing parameter, the no-CMC spacetimes constructed in \cite{CIP05} are both future and past null geodesically incomplete.
\end{theorem}
We do this by using the symmetry of the construction to show that the central cross-section of the gluing region must be both a marginally outer trapped surface (MOTS) and a marginally inner trapped surface (MITS) with particularly rigid geometry. We then use a covering space argument together with Chru{\'s}ciel and Galloway's generalization of the Penrose singularity theorem (Proposition 1.1 of \cite{CG14}). In Section \ref{sec:pre} we recall the localized IMP gluing construction, and in Section \ref{sec:proof} we deduce null incompleteness.

The first and third authors have generalized in \cite{BP19} the results presented here, showing that this incompleteness is not an artifact of the symmetry, but rather a consequence of the geometry of the underlying IMP gluing construction. Therefore, the maximal globally hyperbolic evolution of any (IMP) glued initial data sets admitting noncompact covers are causal geodesically incomplete for sufficiently large values of the gluing parameter.

The foundational work of Robert Bartnik plays a crucial role in the mathematics discussed in this paper.  Indeed, his work in general relativity has had a huge influence over the direction of the field, and many of the currently active branches of research have grown out of the seeds that he planted. On a personal level, the third author is grateful to have known Robert since his time as a graduate student at Stanford in the late 1980s, when Robert visited the department from Australia.  It has always been a pleasure to discuss mathematics with him and learn from him.  The authors are honored to dedicate this work to him.

\section{Acknowledgments} \label{sec:ack}

The authors would like to thank Otis Chodosh and Greg Galloway for helpful discussions and suggestions. The first two authors would also like to thank the organizers---Justin Corvino, Lan-Hsuan Huang, and Damin Wu---of the UConn Summer School in Minimal Surfaces, Flows, and Relativity (supported by the NSF award DMS 1452477, the UConn Math Department, and the UConn College of Library Arts and Sciences) for providing a lively and creative space for junior researchers such as ourselves to collaborate on research. This work was also supported by a grant from the Simons Foundation (279720-DP).

\section{Preliminaries} \label{sec:pre}
Here we recall the CIP construction {\cite{CIP05}, which heavily uses  the IMP gluing construction \cite{IMP02}, to set up
notation and review the known properties of these examples. We begin with a vacuum initial data set $(\TT^{3},\gamma,K)$ which has no global Killing Initial Data (KIDs) and such that for some $p\in\TT^{3}$, there is a neighborhood of $p$ on which $\tau:=\tr_{\gamma}K\equiv0$. 

To achieve this initial setup, one uses of the work of Beig, Chru\'sciel and Schoen \cite{BCS} on the generic absence of KIDs in initial data sets, as well as the work of
Bartnik \cite{bartnik:variational} on the Plateau problem
for prescribed mean curvature spacelike hypersurfaces in a Lorentzian
manifold.

The construction then proceeds by applying a localized form of IMP gluing (see \cite{CD03}) to form a connected sum of  $(\TT^{3},\gamma,K)$ and $(\TT^{3},\gamma,-K)$ around the points $p$. Recall that the gluing procedure consists of the following steps:
\begin{itemize}

\item On each copy, we consider the decomposition $\gamma|_{B_{2R}(p)}=dr^{2}+r^{2}h(r)$ in normal coordinates around $p$, where $r$ is the geodesic distance from $p$, and $h$ is a smooth family of metrics on $\SS^{2}$ with $h(0)\equiv\round$, the standard round metric on the unit sphere. In these coordinates, consider the conformal factor:
\begin{equation*}
\psi_{c}(p)=\begin{cases}
1 & p\in\TT^{3}\sm B_{2R}(p) \\
\text{interpolation} & p\in B_{2R}(p)\sm B_{R}(p) \\
r^{1/2} & p\in B_{R}(p)
\end{cases},
\end{equation*}
where by interpolation, we here and henceforth mean interpolation of the explicitly defined functions using radial cutoff functions with bounded derivatives. Now blow up by $\psi_{c}^{-4}$ so that $\gamma$ approaches the cylindrical metric as $r\searrow0$. That is, for $t=-\log r$, the metric $\gamma_{c}=\psi_{c}^{-4}\gamma$ decomposes  as $\gamma_{c}|_{B_{R}(p)\sm\{p\}}=dt^{2}+h(e^{-t})$.

\item The cylinders are cut off at the parameter $t=T$ for $T$ large, and, distinguishing data on the two $\TT^{3}$'s by the subscripts 1 and 2, we glue by the rule: $(t_{1},\theta_{1})\sim(t_{2},\theta_{2})$ if $t_{2}=T-2\log(R)-t_{1}$ and $\theta_{2}=-\theta_{1}$. On the manifold $M\approx\TT^{3}\#\TT^{3}$, we define a new coordinate $s$, on the glued cylinder  (denoted by $C_{T}$) by
\begin{equation*}
s=t_{1}-\log(R)-T/2=T/2-\log(R)-t_{2}.
\end{equation*}
New data are then constructed by cutoff functions as follows:
\begin{equation*}
\gamma_{T}:=\chi_{1}\gamma_{1}+\chi_{2}\gamma_{2}\quad\text{and}\quad \mu_{T}:=\chi_{1}\mu_{1}+\chi_{2}\mu_{2},
\end{equation*}
where $\mu_{1} $ and $\mu_{2}$ are the transverse-traceless parts of $K$ and $-K$, respectively, and $\{\chi_{1},\chi_{2}\}$ is a partition of unity with respect to an open cover of $M$ whose intersection consists of $\{(s,\theta)\in C_{T}\,:\,s\in(-1,1)\}$. Note that the above definitions also yield a new second fundamental form $K_{T}:=\chi_{1}K_{1}+\chi_{2}K_{2}$. We also define a new conformal factor
\begin{equation*}
\psi_{T}=\wt{\chi}_{1}\psi_{1}+\wt{\chi}_{2}\psi_{2},
\end{equation*}
where $\wt{\chi}_{1},\wt{\chi}_{2}$ are cutoff functions such that $\wt{\chi}_{i}|_{\TT^{3}_{j}\sm B_{R}(p_{j})}=\delta_{ij}$ and on $C_{T}$,
\begin{align*}
\wt{\chi}_{1}(s)&=\begin{cases}
1 & s\in[-T/2,T/2-1) \\
\text{interpolation} & s\in[T/2-1,T/2-1/2) \\
0 & s\in[T/2-1/2,T/2]
\end{cases},\quad\text{and} \\
\wt{\chi}_{2}(s)&=\begin{cases}
0 & s\in[-T/2,1/2-T/2] \\
\text{interpolation} & s\in(1/2-T/2,1-T/2] \\
1 & s\in(1-T/2,T/2]
\end{cases}.
\end{align*}
\end{itemize}

\begin{remark} \label{rem:sym}
The cutoff functions must be chosen so that $M$ satisfies the following symmetry:
\begin{enumerate}
\item There exists a diffeomorphism $\beta:M\to M$ that takes a point on one $\TT^{3}$ to the corresponding one on the other $\TT^{3}$. In particular, on the gluing neck, $\beta(s,\theta)=(-s,\theta)$, so the cross-section $s=0$ is fixed by $\beta$.
\item The reflection $\beta$ satisfies: $\beta^{*}\gamma_{T}=\gamma_{T}$ and $\beta^{*}K_{T}=-K_{T}$.
\end{enumerate}
\end{remark}

\begin{remark} For the purposes of this paper, and in accord with the construction in \cite{IMP02}, we call  $T$ as the \textit{gluing parameter}. We expect that the geometry of the central gluing neck of the resulting initial data set behaves like a small perturbation of a neighborhood of the minimal 2-sphere in a time-symmetric slice of the Schwarzschild spacetime with mass $m_T \sim  Ce^{-\alpha T}$ for positive constants $C$ and $\alpha$ which are independent of $T$.  Existence of solutions to the conformally modified momentum and Hamiltonian constraints follows from perturbation arguments as $T\rightarrow \infty$, where we see a degeneration in the geometry of the initial data sets:  
\end{remark}

\begin{itemize}
\item The new tensor $\mu_{T}$ is perturbed by solving a boundary value problem with the vector Laplacian so that the resulting tensor, $\wt{\mu}_{T}:=\mu_{T}-\sigma_{T}$, is transverse-traceless in the gluing region (where $\tau$ is constant) and $\sigma_{T}=0$ outside the gluing region---since $\wt{\mu}_{T}$ is only different from $\mu_{1}$ and $\mu_{2}$ in the very center of the neck where the latter two are interpolated, the perturbation is only needed in that area. All analysis is done in function spaces so that the above symmetry is preserved.

\item Using a contraction mapping argument, the conformal factor $\psi_{T}$ is perturbed so that the resulting function, $\wt{\psi}_{T}:=\psi_{T}+\eta_{T}$, satisfies the Lichnerowicz equation with boundary conditions fixing $\eta_{T}$ to be 0 outside of the gluing region. Again, all analysis is done in function spaces that preserve the reflection symmetry. In addition, the perturbation term $\eta_{T}$ satisfies the following weighted H{\"o}lder estimate:
\begin{equation} \label{eqn:etabd}
||\eta_{T}||_{k+2,\alpha,\delta}:=||w_{T}^{-\delta}\eta_{T}||_{k+2,\alpha}\le Ce^{-T/4},
\end{equation}
where $\delta\in(0,1)$, $w_{T}|_{C_{T}}:=e^{-T/4}\cosh(s/2)$, the unweighted H{\"o}lder norm is defined as in Definition 2 of \cite{IMP02}, and $C>0$ is independent of $T$.

\item Applying the above procedure for small enough $R$, a compactly supported, smooth deformation procedure per \cite{CD03} is applied across annuli about the boundaries of the gluing neighborhoods. This perturbation agrees with the IMP construction near the middle of the gluing neck and the original data near the gluing neighborhood boundaries.

\item The new initial data, given by:
\begin{equation} \label{eqn:glued}
(M,\wt{\gamma}_{T},\wt{K}_{T})=\left(\TT^{3}\#\TT^{3},\,\wt{\psi}_{T}^{4}\gamma_{T},\,\wt{\psi}_{T}^{-2}\wt{\mu}_{T}+{\tr K\over3}\wt{\psi}_{T}^{4}\gamma_{T}\right),
\end{equation}
satisfies the Einstein vacuum constraint equations as well as the symmetry of Remark \ref{rem:sym}. When convenient, we suppress the dependence on $T$ and denote the final initial data set by $(M,\wt{\gamma},\wt{K})$. However, for the above analysis as well as the bounds we use below, it is necessary that $T$ be sufficiently large.
\end{itemize}

\section{Proof of the Theorem} \label{sec:proof}

Let $\wt{\Sigma}$ be the cross-section $\{(s,\theta)\in C_{T}:\,s=0\}$ with data induced by $(M,\wt{\gamma},\wt{K})$. Using the symmetry of Remark \ref{rem:sym}, we demonstrate below that $\wt{\Sigma}$ is a MOTS and a MITS. First note that
\begin{equation*}
\wt{K}|_{\wt{\Sigma}}=\beta^{*}\wt{K}|_{\wt{\Sigma}}=-\wt{K}|_{\wt{\Sigma}},
\end{equation*}
so $\wt{K}|_{\wt{\Sigma}}\equiv0$. Likewise, if we let $\nu$ be the unit normal pointing in the positive $s$ direction (since orthogonality properties of the original metric are preserved under conformal transformations, the unit normal to $\wt{\Sigma}$ after the final conformal transformation is a rescaling of ${\partial\over\partial s}$), and if we let $H_{+}$ and $H_{-}$ be the mean curvatures of $\wt{\Sigma}$ associated to $\nu$ and $-\nu$, respectively, we have
\begin{equation*}
H_{+}=\beta^{*}H_{+}=H_{-}=-H_{+},
\end{equation*}
so $H_{\wt{\Sigma}}\equiv0$. Thus,  $\wt{\Sigma}$ is a spacetime minimal surface, and in particular satisfies the MOTS equation
\begin{equation*}
\tr_{\wt{\Sigma}}K+H_{\wt{\Sigma}}=0.
\end{equation*}

Now consider the following covering space of $M$: given one of the tori at the beginning of the gluing construction, take a universal cover 
and on each copy of the gluing neighborhood, identically glue in the other torus (using pullback data on the universal cover of the first torus) as described above. Call the resulting space $\cal{N}$. Fixing a single copy of $\wt{\Sigma}$ in this covering space, we are in the situation of Proposition 1.1 of \cite{CG14}:
\begin{enumerate}[(i)]
\item Because the CIP construction is vacuum, it trivially satisfies the null energy condition, and $\cal{N}$ is a noncompact Cauchy surface for its spacetime evolution.
\item The hypersurface $\wt{\Sigma}$ is a closed, connected MOTS, and its complement in $\cal{N}$ consists of two disjoint open sets, one of which has noncompact closure (without loss of generality, let $\nu$ point toward this end).
\item We must show that either the null second fundamental form $\chi$ of $\wt{\Sigma}$ is not identically zero, or that $\wt{\Sigma}$ is strictly stable.
\end{enumerate}
We now show that the last item is satisfied.

Suppose that $\chi\equiv0$. It suffices to show that there exists a function $\phi\in C^{\infty}(\wt{\Sigma})$ such that $L\phi>0$, where $L:C^{\infty}(\wt{\Sigma})\to C^{\infty}(\wt{\Sigma})$ is the MOTS stability operator (see \cite{AMS05}):
\begin{align*}
L\phi:&=-\Delta\phi+2\langle X,\nabla\phi\rangle \\
&\quad+\left({1\over2}R_{\wt{\Sigma}}-(\mu+J(\nu))-{1\over2}|\chi|^{2}+\div X-|X|^{2}\right)\phi,
\end{align*}
and where all differential operators and inner products are taken with respect to the induced metric on $\wt{\Sigma}$, $R_{\wt{\Sigma}}$ is the scalar curvature of $\wt{\Sigma}$, $\mu$ and $J$ are the respective energy and momentum densities, and $X:=\left(\wt{K}(\nu,\cdot)|_{T\wt{\Sigma}}\right)^{\sharp}$. Now in our case, $\mu$ and $J$ are both zero because $\cal{N}$ is vacuum, and $\chi$ disappears by assumption. In addition, since $\wt{K}|_{\wt{\Sigma}}\equiv0$ and all derivatives are taken with respect to $\wt{\Sigma}$, all $X$ terms disappear, whence the stability operator simplifies to
\begin{equation*}
L\phi=-\Delta\phi+{1\over2}R_{\wt{\Sigma}}\phi.
\end{equation*}

Let $\phi\equiv1$, so we are left to show that $R_{\wt{\Sigma}}>0$. But now recall that before the final conformal transformation in the IMP gluing construction, the spherical cross sections close to the middle of the neck have metrics that are arbitrarily close to the standard spherical metric for $T$ large. Thus, we may choose $T$ large enough so that the scalar curvature of the $s=0$ slice is positive. Denote the $s=0$ slice prior to the final conformal transformation $(\Sigma,h)$---that is, the data on $\Sigma$ is induced by $(M,\gamma_{T},K_{T})$. Then using the formula for scalar curvature after a conformal transformation, we see the scalar curvature of $\wt{\Sigma}$ is given by:
\begin{align*}
R_{\wt{\Sigma}}&=(\wt{\psi}_{T})^{-4}\left(R_{\Sigma}-4\Delta(\log(\wt{\psi}_{T}))\right) \\
&=(\wt{\psi}_{T})^{-4}\left(R_{\Sigma}-4\left({||\nabla\eta_{T}||^{2}\over\wt{\psi}_{T}^{2}}-{\Delta\eta_{T}\over\wt{\psi}_{T}}\right)\right),
\end{align*}
where all derivatives and inner products are taken with respect to the induced metric on $\Sigma$. Note that the second equality follows because $\psi_{T}|_{C_{T}}$ is a function of $s$, so it is constant on $\Sigma$. We must show that the last two terms can be made arbitrarily small for $T$ large. From our definitions of $\psi_{T}$ and $w_{T}$ in Section \ref{sec:pre}, we see that
\begin{equation*}
\psi_{T}|_{\Sigma}=\psi_{T}(0)=2e^{-T/4}\quad\text{and}\quad w_{T}|_{\Sigma}=e^{-T/4}.
\end{equation*}
Combining these with the bound on $\eta_{T}$ in (\ref{eqn:etabd}), we obtain
\begin{equation*}
||\eta_{T}||_{k+2}^{\Sigma}\lesssim e^{-(1+\delta)T/4},
\end{equation*}
which yields
\begin{equation*}
||\eta_{T}||_{k+2}^{\Sigma}\lesssim e^{-\delta T/4}\psi_{T}|_{\Sigma}
\end{equation*}
for $\delta\in(0,1)$. Thus, for $T$ sufficiently large, we indeed have $||\nabla\eta_{T}||^{2}$ and $|\Delta\eta_{T}|$ are negligible compared with $\wt{\psi}_{T}^{2}$ and $\wt{\psi}_{T}$, respectively, and hence  $R_{\wt{\Sigma}}$ is positive as desired. Thus, if $\chi\equiv0$, we have $\wt{\Sigma}$ is a strictly stable MOTS, so (iii) is satisfied, and we conclude that any spacetime evolution of $\cal{N}$ is future null geodesically incomplete.

It remains to show that any spacetime evolution $\wh{M}$ of $M$ is future null geodesically incomplete. We use the following lemma from \cite{GL18b}:
\begin{lemma}
Let $(M,\wt{\gamma},\wt{K})$ be a smooth spacelike Cauchy surface in a spacetime $(\wh{M},g)$, and suppose $\pi:\cal{N}\to M$ is a Riemannian covering map. Then there exists a Lorentzian covering map $\wh{\pi}:\wh{\cal{N}}\to\wh{M}$ extending $\pi$ such that $(\cal{N},\pi^{*}\wt{\gamma},\pi^{*}\wt{K})$ is a Cauchy surface for the spacetime $\wh{\cal{N}}$.
\end{lemma}

\begin{center}
\begin{tikzcd}[ampersand replacement=\&]
(\cal{N},\pi^{*}\wt{\gamma},\pi^{*}\wt{K})\arrow{d}{\pi} \arrow[hook]{r} \& (\wh{\cal{N}},\wh{\pi}^{*}g)\arrow{d}{\wh{\pi}} \\
(M,\wt{\gamma},\wt{K})\arrow[hook]{r} \& (\wh{M},g)
\end{tikzcd}
\end{center}

Now suppose $\wh{M}$ is future null geodesically complete. Since $\wh{\cal{N}}$ is future null geodesically incomplete, there exists a future inextendible smooth null geodesic $\zeta:[0,\alpha)\to\wh{\cal{N}}$ that terminates at affine parameter $\alpha<\infty$. Consider the smooth null geodesic $\wh{\pi}(\zeta)\subset\wh{M}$. Then by future null completeness of $\wh{M}$, we must be able to find a smooth null geodesic $\wh{\zeta}:[0,\infty)\to\wh{M}$ extending $\wh{\pi}(\zeta)$. Let $\veps>0$ be small enough so that $\wh{\zeta}(\alpha-\veps,\alpha+\veps)$ is contained in a single evenly-covered neighborhood $U\subset\wh{M}$, and pick a smooth local section $\sigma:U\to\wh{\cal{N}}$ of the covering such that $\zeta(t)=\sigma\circ\wh{\pi}\circ\zeta(t)$ for every $t\in(\alpha-\veps,\alpha)$. Then we see that the smooth null geodesic $\xi:[0,\alpha+\veps)\to\wh{\cal{N}}$ defined by
\begin{equation*}
\xi(t):=\begin{cases}
\zeta(t) & t\in[0,\alpha) \\
\sigma\circ\wh{\zeta}(t) & t\in(\alpha-\veps,\alpha+\veps)
\end{cases}
\end{equation*}
extends $\zeta$ to the future, which contradicts our assumption on $\zeta$. Thus, we must indeed have that any spacetime evolution of $M$ is future null geodesically incomplete.

Lastly, note that since $\wt{\Sigma}$  is also marginally inner trapped we may take a time reversal of the above argument---using a covering space that ``unwraps" the other torus---to conclude past null geodesic incompleteness.

\qed

\end{document}

%% file: MyDefns.tex
\def\veps{\varepsilon}

\def\sm{\setminus}

\def\SS{{\mathbb S}}
\def\TT{{\mathbb T}}

\def\1{{\mathbbm 1}}

\def\cal{\mathcal}

\def\tr{\operatorname{tr}}

\def\round{{\stackrel{\circ}{g}}} 

\def\div{\operatorname{div}}

\def\wt{\widetilde}

\def\wh{\widehat}